\pdfoutput=1

\documentclass[12PT]{iopart}
\usepackage[pdftex]{graphicx}

\begin{document}


\title{$\Upsilon$ MEASUREMENT IN STAR}

\author{\footnotesize MAURO R. COSENTINO for the STAR COLLABORATION}

\address{Instituto de F\'{i}sica, University of S\~ao Paulo, Rua do Mat\~ao, 187 - travessa R S\~ao Paulo, SP 05508-090,Brazil}

\ead{mcosent@dfn.if.usp.br}


\begin{abstract}

We present preliminary results of $\Upsilon$ production in p+p collisions
at $\sqrt{s}=$200 GeV at central rapidity. This measurement was
performed at the STAR experiment through the $\Upsilon\rightarrow 
e^+e^-$ decay channel. In this manuscript we describe the experimental 
details, from the development of a specially designed trigger setup 
to the analysis methods used to discriminate electrons from hadrons. 
The production cross-section obtained 
$B\times\frac{d\sigma}{dy}\vert_{y=0}=$91$\pm$28$(\textit{stat.})\pm$22$(\textit{sys.)}$ pb is compatible with our 
expectations based on pQCD calculations.

\end{abstract}

\section{Introduction}

 The excited states of heavy-quarkonium are predicted to be 
strongly modified in 
heavy-ion collisions with respect to p+p \cite{1,2}. Therefore a full 
spectroscopy of the $\Upsilon$ states is good way to observe any 
suppression pattern that corroborates to the existence of deconfinement 
and the specific pattern of suppression can provide 
experimentalists with a thermometer of the QGP.

\section{Experimental Setup}

STAR \cite{3, 4, 5} is a large acceptance, multi purpose experiment composed 
of many individual detector subsystems installed inside a large 
solenoidal 
magnet capable of creating uniform magnetic fields up to 0.5 Tesla. In 
the following, we describe the detectors which are relevant to the present 
analysis.
The Time Projection Chamber (TPC) \cite{6} has a pseudorapidity coverage of 
$\vert\eta\vert<$ 1.8  for collisions in the center of STAR with full azimuthal 
coverage. For charged tracks in the acceptance, the TPC provides up to 45 
independent spatial and specific ionization energy loss ($dE/dx$) measurements. The $dE/dx$ 
measurement, in combination with the momentum measurement, determines the 
particle mass within a limited kinematic region.
The Barrel Electromagnetic Calorimeter (BEMC)\cite{7}, located just inside 
the coils 
of the STAR solenoidal magnet with a radius relative to beam line of 2.3 m, 
is a lead-scintillator sampling electromagnetic calorimeter with equal volumes 
of lead and scintillator. The electromagnetic energy resolution of the detector 
is $\frac{dE}{E}\sim\frac{16\%}{\sqrt{E}}$. The results presented in 
this 
work used for the first time the full coverage of the BEMC in the 2006 RHIC run, 
consisting of 120 modules,  with coverage from $\vert\eta\vert<1$ and 
full azimuthal coverage. Each BEMC module is composed of 40 towers (20 towers 
in $\eta$ by 2 towers in $\phi$) constructed to project to the center of the 
STAR detector. The tower size is $(\Delta\eta,\Delta\phi) = (0.05, 0.05)$. 
The tower depth is 21 radiation lengths (X$_0$), corresponding to a little 
less than 1 hadronic interaction length.

  \subsection{Trigger Setup}

  The trigger looks for the $\Upsilon\rightarrow e^++e^-$ decay channel and it is a 
  two-level setup, where the high invariant mass allows the trigger to 
  be simpler than other two-level triggers \cite{8}.  The L0 and L2 triggers 
  for the $\Upsilon$ are detailed below.

    \subsubsection{L0 Trigger}

    The L0 trigger is a fast hardware trigger, taking a decision for 
    every bunch crossing. It is simply is a high-tower trigger, which
    consists of at least one BEMC tower with energy above the threshold of 3.5 GeV.
    When such an event happens it starts the L2 trigger.

    \subsubsection{L2 Trigger}

    The L2 trigger is a software level decision. It finds towers 
    with energy similar to the L0 threshold and uses them as 
    seeds to the L2 clusters, that are groups of 3 towers made 
    by the seed plus the 2 surrounding towers with the highest energy. 
    As 
    the decay channel is  $e^++e^-$, the L2 algorithm  takes pairs of 
    clusters and calculates the invariant mass of the cluster pair using 
    the formula
    \begin{eqnarray}
      M^2 = 2E_1E_2(1-cos\theta_{12})\,
      \label{eq:mass2}
    \end{eqnarray}
    where $E_i$ is the energy of the cluster $i$ (1 or 2) and $\theta_{12}$ 
    is obtained as the angle formed by the two straight lines that go 
    from the vertex to the cluster positions. The vertex is assumed to be 
    at (0,0,0). The implemented $\Upsilon$ trigger has a high efficiency, 
    above 80$\%$ and a large rejection power ($\sim$100 for each of the 2 levels, 
    bringing it up to $\sim$$10^4$). 
    Due to the high invariant mass for which it is designed to work $-$ where 
    background is much lower $-$, the $\Upsilon$ trigger is efficient on both, 
    p+p multiplicities up to central Au+Au. This trigger also exploits 
    the full STAR acceptance 
    of 2$\pi$ in $\phi$ and $\vert\eta\vert<1$.

\section{Electron Identification}

The process of electron identification using the STAR barrel calorimeter 
is based on a pre-selection of electron candidates from the TPC $dE/dx$ 
measurement. Figure 1 (left) shows a typical TPC $dE/dx$ distribution 
for tracks where the number of points used to calculate the $dE/dx$ is greater 
than 20 and the momenta of the tracks $p>$1 GeV/c. It is possible to see that 
electrons can be pre-selected by the range 3$<dE/dx<$5 keV/cm.

\begin{figure}[th]
\begin{center}
\includegraphics[width=1.0\textwidth]{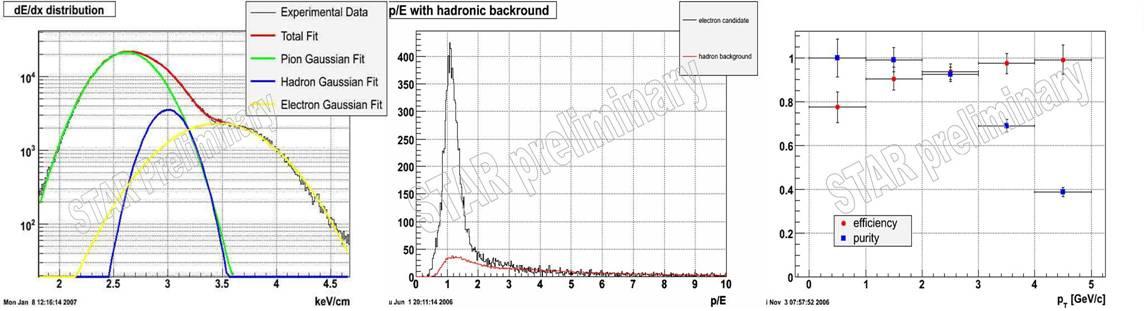}
\vspace{-0.8cm}
\caption{On the  left panel it is presented the $dE/dx$ distribution for all
  charged particles that pass through the TPC. In this panel there are 3 
  gaussian fits, one for electrons (right most), one for pions and the other for the 
  remaining hadrons. The central plot shows the p/E distribution within the 
  electron band and the hadronic background obtained from the hadron dE/dx band 
  (thick line). The right panel shows the efficiency (round dots) and purity 
  (square) of this method as function of $p_T$.}
\end{center}
\end{figure}

After the electron candidates are selected, their tracks are projected on the 
BEMC and the energy deposited in the tower hit by the track is compared 
to its momentum. Due to their small masses electrons deposit all their 
energies on the BEMC towers and thus should have an $p/E_{tower}$ 
$\sim$1. Hadrons 
have a wider distribution of $p/E$. Still on figure 1, the left panel 
shows the $p/E_{tower}$ spectrum for the electron candidates in which 
it is possible to see a well defined electron peak. 
The peak is not centered at one due to energy leakage to surrounding towers. 
The residual hadronic background is shown as a thick line in the spectrum. 
With all this information it is possible to make a series of selected cuts 
that provide a good particle identification (PID) and hadron rejection for 
the desired electrons. The main cuts are $p>$2 GeV/c, 3.4$<dE/dx<$5.0 keV/cm and 
$p/E<$2. These are the set of cuts that give the relation efficiency $\times$ purity 
on the electron identification with best signal-to-background.Some of the 
studies for the PID efficiency are presented on figures 1 (right) and 2. On the 
first one, it is possible to see that the efficiency approaches unity as $p_T$ 
increases, while the purity becomes poorer. This reflects the fact that the 
electron and hadron bands move to higher $dE/dx$ ranges with higher $p_T$.  
Figure 2, on the other hand, shows how the purity and the efficiency 
vary with changes on the cuts parameters. This figure also shows another 
important 
feature that is the hadron rejection power of these cuts.
The expressions used to calculate the efficiency, purity, rejection power 
and normalized rejection power are:

\begin{list}{}{}
  \item{ {\bf Efficiency:} (selected electrons)/(total electrons); }
  \item{ {\bf Purity:} (selected electrons)/(selected electrons+hadrons); }
  \item{ {\bf Rejection power:} (selected electrons)/(selected hadrons); }
  \item{ {\bf Normalized RP:} (probability of selected electrons)/(probability of selected hadrons).}
\end{list}
The number of selected particles is the area of the Gaussian that falls within the cuts. 
The total number of particles is the area under the entire Gaussian. The probability of 
selected particles is the area of the Gaussian that falls within the cuts where the 
Gaussians have been normalized to unity for all species.
 
\begin{figure}[th]
\begin{center}
\includegraphics[width=1.0\textwidth]{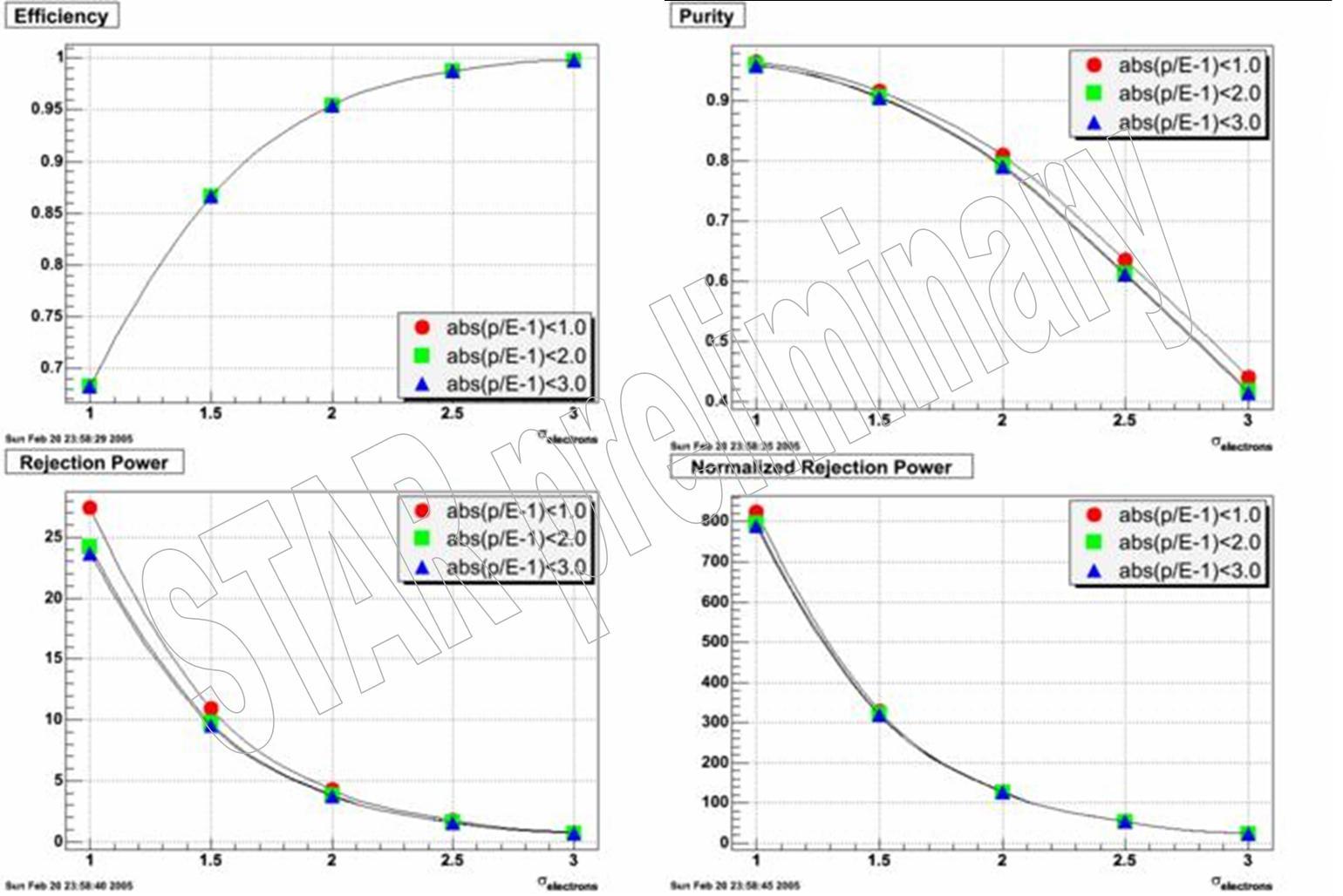}
\vspace{-0.8cm}
\caption{Efficiency, Purity, Rejection Power(RP) and Normalized RP.}
\end{center}
\end{figure}

\section{Results}

After making the electron identification, the electron sample is analyzed 
and the invariant mass between every two electrons are calculated. 
With the results obtained from $e^+e^-$ pairs we calculate the invariant  
mass spectrum for unlike-sign pairs and construct a background spectrum 
using the like-sign pairs($2\sqrt{N^{++}N^{--}}$). 
Figure 3 (left) show $e^+e^-$ combinations of electron candidates
(dots) and the corresponding combinatorial background (line). 
The vertical lines in figure 3 represent the boundaries that ``confine'' the data used to evaluate 
the significance of the signal. The value obtained for it is 3.0$\sigma$.

\begin{figure}[th]
\begin{center}
\includegraphics[width=1.0\textwidth]{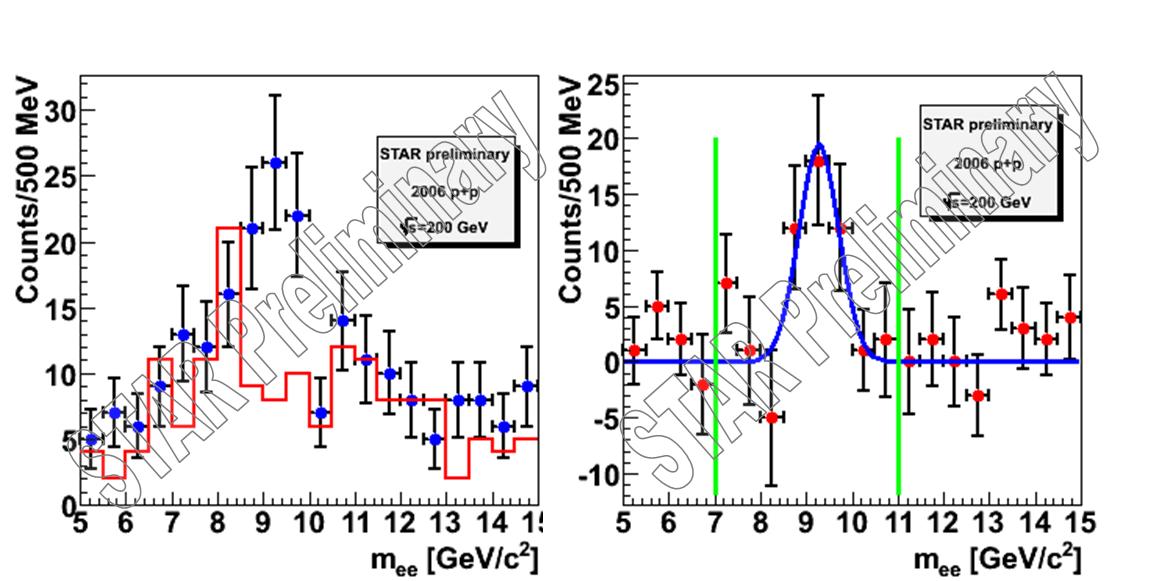}
\end{center}
\vspace{-0.8cm}
\begin{center}
\caption{At left the $\Upsilon$ mass signal constructed by combining $e^+$ and $e^-$ in pairs using 
  expression (2) and its correspondent like-sign combinatorial background. In the right panel
  the net $\Upsilon$ signal after background subtraction.}
\end{center}
\end{figure}

On figure 3 (right) it is possible to see the net $\Upsilon$ signal after  subtraction of combinatorial background. 
Given the magnetic field strength, the expected mass resolution for of 
the $\Upsilon$ peak, obtained from simulations, is about 0.5 GeV/c$^2$, 
consistent with the $\sigma_{gauss}\sim$0.5 GeV/c$^2$ of figure 3. 
The other important remark is that the mean value of the fitted Gaussian is 
shifted down by $\sim$160 MeV/c$^2$, which is due to Bremsstrahlung  effects on 
the electrons when they pass through the inner materials of the STAR experiment 
(like the beam-pipe for example).
With this resolution and the available statistics it is not possible from this 
measurement to discriminate between the 3 S states of the $\Upsilon$ family. 
Considering this year$'$s acceptance, the overall efficiency in the $\Upsilon$ 
measurement and the integrated luminosity it is possible 
to compute the cross section for the  (1S+2S+3S) states using the expression

\begin{eqnarray}
  B\times\frac{d\sigma}{dy}\vert_{y=0} = 
  \frac{N_{\Upsilon}}{dy\times\epsilon_{\Upsilon}\times\int\mathcal{L}dt}\,
  \label{eq:x-section}
\end{eqnarray}

where $\epsilon_{\Upsilon}$,  and $\int\mathcal{L}dt$ are the 
$\Upsilon$ overall efficiency and the 
integrated luminosity, respectively.
With all the value computed we get the cross-section = 91$\pm$28$(\textit{stat.})\pm$22$(\textit{sys.)}$  pb  which 
agrees nicely well with the pQCD-CEM calculation of 91 pb \cite{8,9}.

\section{Perspectives}

For the next year, RHIC will have a long Au+Au run, with much higher statistics, 
and we will be able to use the $\Upsilon$ trigger with all the STAR acceptance for the 
first time in Au+Au. This will be our first opportunity to measure the 
$\Upsilon$ S-states family in a heavy-ion environment. 
This measurement could shed some light on the sequential melting models.

In the mid/near term, RHIC and STAR will receive several upgrades that will enhance 
our measurement capabilities. First, the RHIC upgrade to RHIC-II with e-cooling will 
provide a substantial increase in the luminosity ($\sim$40$\times$RHIC luminosity), which is 
today's big constraint on the $\Upsilon$ measurement. One specific remark is that the collision 
diamond $-$ the longitudinal distribution of beams intersection$-$ is expected to go from today's 
$\sigma$=20 cm to $\sigma$=10 cm what makes the gain in the ``usable luminosity'' larger than the nominal 
increase. On the STAR side, there are also several upgrades that may help the quarkonium program 
and specifically the $\Upsilon$ measurements. The main upgrades are the DAQ1000, which will 
increase considerably the trigger rates giving them enough capabilities to have zero dead time.
Other upgrades that will help the quarkonium program are the full barrel Time-Of-Flight(ToF) 
and a muon detector. The ToF can increase the quality on the particle identification, 
helping to improve electron identification and to increase hadron rejection power. 
It can also help on other quarkonia trigger setups.
With the muon detector it will be possible to make complementary $\Upsilon$ measurements via 
the $\Upsilon\rightarrow \mu^+\mu^-$ decay channel. With all these upgrades, it is estimated 
that the STAR experiment will be able to measure $\sim$11200 $\Upsilon$'s (against 
$\sim$830 max., with RHIC), making the $\Upsilon$ the strength of STAR in the quarkonium program.

\section*{Acknowledgements}

This work was supported in part by the Brookhaven National Laboratory (BNL), by the 
Conselho Nacional de Desenvolvimento Cient\'{\i}fico e Tecnol\'{o}gico (CNPq) and 
by the Coordena\c{c}\~{a}o de Aperfei\c{c}oamento de Pessoal de N\'{\i}vel Superior
(CAPES).

\end{document}